\documentclass{webofc}
\usepackage[varg]{txfonts}   
\def\snn{\sqrt{s_\mathrm{NN}}}
\def\GlauberMUSICUrQMD{\textsc{3d-glauber+music+urqmd}}
\def\iEBEMUSIC{\textsc{iebe-music}}
\def\UrQMD{\textsc{urqmd}}

\begin{document}
\title{The role of longitudinal decorrelations for measurements of anisotropic flow in small collision systems}
%
%

\author{ \firstname{Sangwook} \lastname{Ryu}\inst{1}\fnsep
\and
        \firstname{Bj\"orn} \lastname{Schenke}\inst{2}\fnsep\thanks{\email{bschenke@bnl.gov}} 
        \and
        \firstname{Chun} \lastname{Shen}\inst{1,3}
        \and
        \firstname{Wenbin} \lastname{Zhao}\inst{4}
}

\institute{
           Department of Physics and Astronomy, Wayne State University, Detroit, Michigan 48201, USA
\and
           Physics Department, Brookhaven National Laboratory, Upton, NY 11973, USA
\and
           RIKEN BNL Research Center, Brookhaven National Laboratory, Upton, NY 11973, USA
\and        
           Nuclear Science Division, Lawrence Berkeley National Laboratory, Berkeley, California 94720, USA
          }

\abstract{%
 Within a (3+1)D viscous hydrodynamic model we compute anisotropic flow in small system collisions as performed at the Relativistic Heavy Ion Collider and measured by the STAR and PHENIX Collaborations. We emphasize the importance of the rapidity dependence of the geometry for interpreting the differences encountered in measurements by the two collaborations.
}
\maketitle
\section{Introduction}
\label{sec:intro}
The PHENIX Collaboration performed measurements of anisotropic flow in 
collisions of proton (p), deuteron (d), and helium-3 ($^3$He) projectiles with gold (Au) nuclei to determine if previously observed ``flow-like"  patterns in small system collisions \cite{Dusling:2015gta,Loizides:2016tew,Schlichting:2016sqo,Nagle:2018nvi,Schenke:2019pmk,Schenke:2021mxx} are geometry driven and attributable to the formation of quark-gluon plasma (QGP) droplets. PHENIX data \cite{PHENIX:2018lia} demonstrated hierarchies among the elliptic flow $v_2$ and triangular flow $v_3$ coefficients, a result that was expected if a strongly interacting system is created, based on the differences in initial geometries.  

In contrast to the PHENIX data, results reported by the STAR Collaboration \cite{Lacey:2020ime,STAR:2022pfn} show that the observed $v_3(p_T)$ are system independent. Sub-nucleon fluctuations are highlighted in \cite{Lacey:2020ime,STAR:2022pfn} as a possible explanation for why initial geometries could be more similar than expected from purely nucleon driven fluctuating geometries.

To understand the differences in the results from STAR and PHENIX, it is important to consider the two experiments' use of different pseudo-rapidity ranges in the two-particle correlation measurements that yield the azimuthal anisotropy coefficients. 

While PHENIX measures two-particle correlations between backward rapidity and mid-rapidity regions \cite{PHENIX:2018hho,PHENIX:2018lia}, STAR employs only the mid-rapidity region \cite{Lacey:2020ime,STAR:2022pfn}. As in small asymmetric systems longitudinal decorrelations are significant \cite{CMS:2015xmx} and boost-invariance is broken~\cite{Bozek:2015swa,Schenke:2016ksl,Shen:2020jwv,Jiang:2021ajc,Wu:2021hkv, Zhao:2022ayk}, a full understanding of the RHIC results requires simulations considering the full (3+1)D dynamics and an initial state that describes the fluctuating geometry as a function of not just the transverse plane but also the longitudinal direction.

In this work, we perform calculations of anisotropic flow in small system collisions at $\snn=200\,{\rm GeV}$ using a (3+1)D framework combining a dynamic initial state with MUSIC viscous relativistic hydrodynamics and hadronic transport~\cite{Zhao:2022ugy,Shen:2022oyg}.

\section{Model}
\label{sec:model}
We use the \GlauberMUSICUrQMD{} hybrid  model within the open-source \iEBEMUSIC{} framework \cite{iEBEMUSIC} to study particle production and flow in p+Au, d+Au, and $^3$He+Au collisions at $\snn=200$\,GeV.
The initial state model deposits energy, momentum, and baryon number dynamically into the fluid and obeys exact energy-momentum conservation \cite{Shen:2017bsr, Shen:2022oyg}. Colliding nucleons are determined as in conventional Monte Carlo Glauber models. Their interaction is then implemented by connecting a string between nucleon constituents and having them dynamically decelerate. The rapidity loss is sampled from a phenomenologically motivated distribution that depends on the initial constituent rapidity. The lost energy and momentum are then transferred to the medium via source terms. This method results in events with a fluctuating structure in all three spatial dimensions.

For the fluid system we use the equation of state {\tt NEOS-BQS} \cite{Monnai:2019hkn} employing strangeness neutrality, $n_s=0$, and setting the net electric charge-to-baryon density ratio to $n_Q/n_B=0.4$, a value most appropriate for large nuclei.
The specific shear and bulk viscosities are parametrized as functions of baryon chemical potential and temperature with parameters chosen to reproduce the mean transverse momentum and differential flow in central $^3$He+Au collisions. The  Cooper-Frye particlization of fluid cells is performed on a hyper-surface with a constant energy density of $e_{\rm sw}=$0.50 GeV/fm$^3$ using the Cornelius algorithm \cite{Huovinen:2012is} and the open-source code package \textsc{iSS} \cite{Shen:2014vra, iSS_code}. The produced hadrons are then fed into the hadron cascade model, \UrQMD \cite{Bass:1998ca, Bleicher:1999xi, UrQMD} for further scatterings and decays until kinetic freeze-out is achieved dynamically. 

\begin{figure}
\centering
\includegraphics[width=\textwidth,clip]{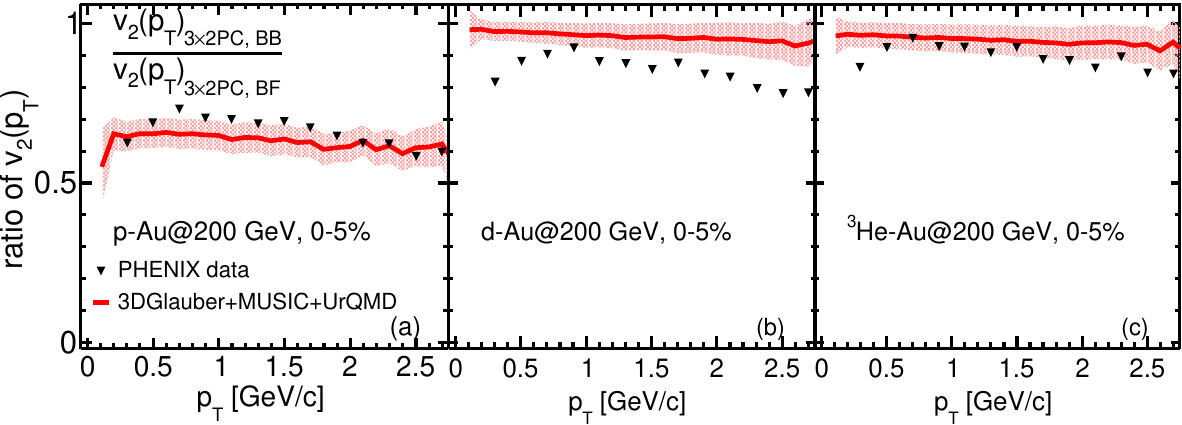}
\caption{The ration of results for $v_2(p_T)$ obtained using two different detector combinations compared to PHENIX data \cite{PHENIX:2021ubk,PHENIX:2022nht}.}
\label{fig2}       
\end{figure}

\section{Results}
Our main results, reported in \cite{Zhao:2022ugy}, showed that a significant part of the differences between STAR and PHENIX measurements can be explained once the longitudinal structure is included. Here, we support the finding of the importance of longitudinal structure with additional results from our analysis that have not been shown in \cite{Zhao:2022ugy}. 

First, Fig.\,\ref{fig2} shows the ratio of results for $v_2(p_T)$ using the 3×2PC method for two different detector combinations in PHENIX, FVTXS-CNT-FVTXN (backward-forward, BF) and BBCS-FVTXS-CNT (backward-backward, BB) from our calculation, compared to the PHENIX data\cite{PHENIX:2021ubk,PHENIX:2022nht}. The trends are well described across the different systems with very good agreement in p+Au and $^3$He+Au collisions.\footnote{We note that both $v_2(p_T)_{3\times2PC,BB}$ and $v_2(p_T)_{3\times2PC,BF}$ in p+Au collisions are underestimated by the model \cite{Zhao:2022ugy}.} This is another demonstration of the fact that the longitudinal decorrelations included in our model play an essential role in generating the differences between $v_2(p_T)$ measurements that use detectors at different rapidities.

\begin{figure}
\centering
\includegraphics[width=\textwidth,clip]{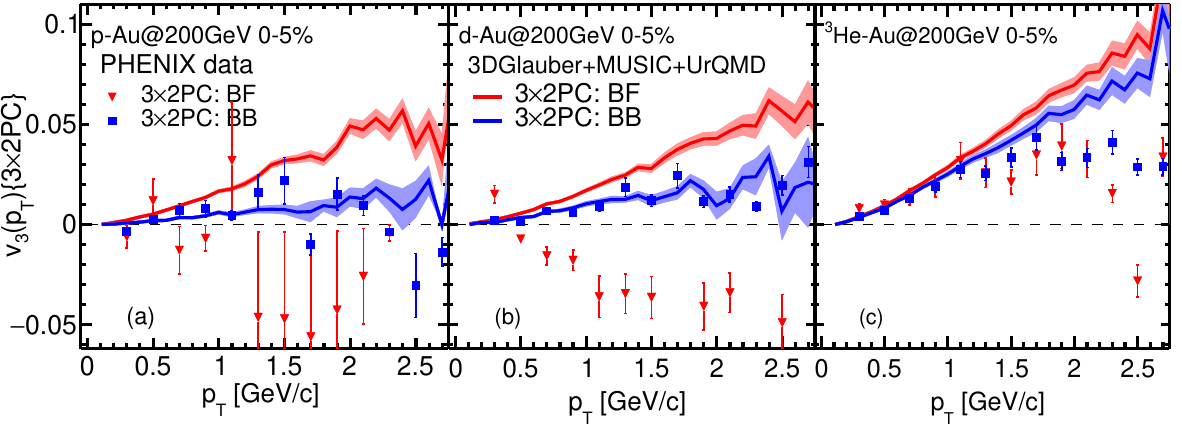}
\caption{The differential triangular flow coefficient $v_3(p_T)$ for charged hadrons using two different methods in three different collision systems compared to PHENIX data \cite{PHENIX:2021ubk,PHENIX:2022nht}.}
\label{fig1}       
\end{figure}

The differential triangular flow coefficients from the BB and BF detector combinations are shown in Fig.\,\ref{fig1}. We find good agreement with the BB data, particularly at $p_T<2\,{\rm GeV}$, using mostly backward (Au-going) detectors. In the case of the backward-forward (BF) detector combination, the experiment finds imaginary results (indicated by negative values in the plot) for p+Au and d+Au collision systems, most likely due to a dominance of non-flow, as the dilute proton or deuteron going direction is included. 

\section{Conclusions}
We have shown additional support for an earlier finding \cite{Zhao:2022ugy} that comparisons of hydrodynamic model calculations to small system anisotropic flow data require the consideration of the produced matter's longitudinal structure. The measured ratio of elliptic flow results from PHENIX backward-backward to forward-backward detector combinations is well described for p+Au and $^3$He+Au systems. 
For triangular flow no strong conclusion can be drawn as the experimental data is non-flow dominated in the two smallest systems, when the forward (small projectile going) direction is included in the analysis.

\section*{Acknowledgments}
W.B.Z. was supported by the National Science Foundation (NSF) under grant numbers ACI-2004571 within the framework of the XSCAPE project of the JETSCAPE collaboration. W.B.Z is also supported by the US DOE under Contract No. DE-AC02-05CH11231, and within the framework of the SURGE Topical Theory Collaboration.
S.R. is supported by the U.S. Department of Energy(DOE) under grant number DE-SC0013460. B.P.S. and C.S. are supported by the U.S. Department of Energy, Office of Science, Office of Nuclear Physics, under DOE Contract No.\,DE-SC0012704 and Award No. DE-SC0021969, respectively. C.S. acknowledges support from a DOE Office of Science Early Career Award. 
This work is in part supported within the framework of the Beam Energy Scan Theory (BEST) Topical Collaboration and under contract number DE-SC0013460.
This research was done using resources provided by the Open Science Grid (OSG) \cite{Pordes:2007zzb, Sfiligoi:2009cct}, which is supported by the National Science Foundation award \#2030508.

\bibliography{references}

\begin{thebibliography}{32}

\bibitem{Dusling:2015gta}
K.~Dusling, W.~Li, B.~Schenke, Int. J. Mod. Phys. E \textbf{25}, 1630002 (2016), \texttt{1509.07939}

\bibitem{Loizides:2016tew}
C.~Loizides, Nucl. Phys. A \textbf{956}, 200 (2016), \texttt{1602.09138}

\bibitem{Schlichting:2016sqo}
S.~Schlichting, P.~Tribedy, Adv. High Energy Phys. \textbf{2016}, 8460349 (2016), \texttt{1611.00329}

\bibitem{Nagle:2018nvi}
J.L. Nagle, W.A. Zajc, Ann. Rev. Nucl. Part. Sci. \textbf{68}, 211 (2018), \texttt{1801.03477}

\bibitem{Schenke:2019pmk}
B.~Schenke, C.~Shen, P.~Tribedy, Phys. Lett. B \textbf{803}, 135322 (2020), \texttt{1908.06212}

\bibitem{Schenke:2021mxx}
B.~Schenke, Rept. Prog. Phys. \textbf{84}, 082301 (2021), \texttt{2102.11189}

\bibitem{PHENIX:2018lia}
C.~Aidala et~al. (PHENIX), Nature Phys. \textbf{15}, 214 (2019), \texttt{1805.02973}

\bibitem{Lacey:2020ime}
R.A. Lacey (STAR), Nucl. Phys. A \textbf{1005}, 122041 (2021), \texttt{2002.11889}

\bibitem{STAR:2022pfn}
 (2022), \texttt{2210.11352}

\bibitem{PHENIX:2018hho}
A.~Adare et~al. (PHENIX), Phys. Rev. Lett. \textbf{121}, 222301 (2018), \texttt{1807.11928}

\bibitem{CMS:2015xmx}
V.~Khachatryan et~al. (CMS), Phys. Rev. C \textbf{92}, 034911 (2015), \texttt{1503.01692}

\bibitem{Bozek:2015swa}
P.~Bozek, A.~Bzdak, G.L. Ma, Phys. Lett. B \textbf{748}, 301 (2015), \texttt{1503.03655}

\bibitem{Schenke:2016ksl}
B.~Schenke, S.~Schlichting, Phys. Rev. C \textbf{94}, 044907 (2016), \texttt{1605.07158}

\bibitem{Shen:2020jwv}
C.~Shen, S.~Alzhrani, Phys. Rev. C \textbf{102}, 014909 (2020), \texttt{2003.05852}

\bibitem{Jiang:2021ajc}
Z.F. Jiang, S.~Cao, X.Y. Wu, C.B. Yang, B.W. Zhang, Phys. Rev. C \textbf{105}, 034901 (2022), \texttt{2112.01916}

\bibitem{Wu:2021hkv}
X.Y. Wu, G.Y. Qin (2021), \texttt{2109.03512}

\bibitem{Zhao:2022ayk}
W.~Zhao, C.~Shen, B.~Schenke, Phys. Rev. Lett. \textbf{129}, 252302 (2022), \texttt{2203.06094}

\bibitem{Zhao:2022ugy}
W.~Zhao, S.~Ryu, C.~Shen, B.~Schenke, Phys. Rev. C \textbf{107}, 014904 (2023), \texttt{2211.16376}

\bibitem{Shen:2022oyg}
C.~Shen, B.~Schenke, Phys. Rev. C \textbf{105}, 064905 (2022), \texttt{2203.04685}

\bibitem{iEBEMUSIC}
We us v0.5 of the iEBE-MUSIC framework, which can be downloaded from \url{https://github.com/chunshen1987/iEBE-MUSIC}.

\bibitem{Shen:2017bsr}
C.~Shen, B.~Schenke, Phys. Rev. C \textbf{97}, 024907 (2018), \texttt{1710.00881}

\bibitem{Monnai:2019hkn}
A.~Monnai, B.~Schenke, C.~Shen, Phys. Rev. C \textbf{100}, 024907 (2019), \texttt{1902.05095}

\bibitem{Huovinen:2012is}
P.~Huovinen, H.~Petersen, Eur. Phys. J. A \textbf{48}, 171 (2012), \texttt{1206.3371}

\bibitem{Shen:2014vra}
C.~Shen, Z.~Qiu, H.~Song, J.~Bernhard, S.~Bass, U.~Heinz, Comput. Phys. Commun. \textbf{199}, 61 (2016), \texttt{1409.8164}

\bibitem{iSS_code}
The iSS code package is an open-source particle sampler based on the Cooper-Frye freeze-out prescription. It converts fluid cells to particle samples. This work uses v1.0 of the iSS, which can be downloaded from \url{https://github.com/chunshen1987/iSS/releases}.

\bibitem{Bass:1998ca}
S.A. Bass et~al., Prog. Part. Nucl. Phys. \textbf{41}, 255 (1998), \texttt{nucl-th/9803035}

\bibitem{Bleicher:1999xi}
M.~Bleicher et~al., J. Phys. G \textbf{25}, 1859 (1999), \texttt{hep-ph/9909407}

\bibitem{UrQMD}
We use the official UrQMD v3.4 and set it up to run as the afterburner mode, \url{https://bitbucket.org/ Chunshen1987/urqmd_afterburner/src/master/}.

\bibitem{PHENIX:2021ubk}
U.A. Acharya et~al. (PHENIX), Phys. Rev. C \textbf{105}, 024901 (2022), \texttt{2107.06634}

\bibitem{PHENIX:2022nht}
U.A. Acharya et~al. (PHENIX) (2022), \texttt{2203.09894}

\bibitem{Pordes:2007zzb}
R.~Pordes et~al., J. Phys. Conf. Ser. \textbf{78}, 012057 (2007)

\bibitem{Sfiligoi:2009cct}
I.~Sfiligoi, D.C. Bradley, B.~Holzman, P.~Mhashilkar, S.~Padhi, F.~Wurthwrin, WRI World Congress \textbf{2}, 428 (2009)

\end{thebibliography}


\end{document}